\documentclass[
a4paper,%
aps,%
12pt,%
final,%
notitlepage,%
oneside,%
onecolumn,%
nobibnotes,%
nofootinbib,%
superscriptaddress,%
noshowpacs]%
{revtex4}
\usepackage[utf8]{inputenc}
\usepackage[T1]{fontenc}
\usepackage{amsfonts}
\usepackage{amssymb}
\usepackage{amsmath}
\usepackage{accents}
\usepackage{graphicx}
\usepackage{color}
\usepackage[colorlinks,hyperindex,plainpages=false]{hyperref}
\allowdisplaybreaks

\newcommand{\dd}{\mathrm{d}}
\newcommand{\lc}[1]{\accentset{\circ}{#1}}

\begin{document}

\noindent {\it Astronomy Reports, 2026, Vol. , No. }
\bigskip\bigskip  \hrule\smallskip\hrule
\vspace{35mm}


\title{Dynamical systems in quadratic teleparallel cosmology\footnote{Paper presented at the Sixth Zeldovich meeting, an international conference in honor of Ya. B. Zeldovich held in Pescara, Italy on July 13--17, 2026. Published by the recommendation of the special editors: R. Ruffini, N. Sahakyan and G. V. Vereshchagin.}}

\author{\bf \copyright $\:$  2026.
\quad \firstname{M.}~\surname{Hohmann}}%
\email{manuel.hohmann@ut.ee}
\affiliation{Laboratory of Theoretical Physics, Institute of Physics, University of Tartu, W. Ostwaldi 1, 50411 Tartu, Estonia}%

\begin{abstract}
\centerline{\footnotesize Received: ;$\;$
Revised: ;$\;$ Accepted: .}\bigskip\bigskip\bigskip

We discuss the cosmological dynamics of general teleparallel quadratic gravity. Using a suitable set of variables, we write the cosmological field equations as a homogeneous dynamical system. We find that this system generically features projective fixed points, where the dynamics can be solved analytically, and exhibits finite time singularities as well as an asymptotically vanishing cosmological expansion.
\end{abstract}

\maketitle

\section{Introduction}
Despite its success describing gravity over a large range of scales~\cite{Baker:2014zba}, General Relativity is challenged by cosmological observations such as the Hubble tension~\cite{DiValentino:2021izs}. This motivates the study of various modified gravity theories~\cite{CANTATA:2021asi,CosmoVerseNetwork:2025alb}. A large class of theories of current interest are teleparallel gravity theories, in which gravity is mediated by the torsion or nonmetricity of a flat connection, instead of the curvature of the Levi-Civita connection~\cite{Hohmann:2022mlc}. In the general teleparallel class of theories, both torsion and nonmetricity are allowed to be non-vanishing. A particularly appealing subclass of these theories is given by general teleparallel quadratic gravity~\cite{BeltranJimenez:2019odq}. The action of this class of theories is quadratic in torsion and nonmetricity, and hence reminiscent of a Yang-Mills theory.

In this article, we study the cosmological dynamics of general teleparallel quadratic gravity, which has been derived in~\cite{Heisenberg:2022mbo}. By choosing a suitable set of dynamical variables, we transform the cosmological field equations into the form of a dynamical system. We then study selected generic properties of this system, in particular the appearance of projective fixed points as well as the asymptotic behavior.

\section{General teleparallel cosmology}
In this article we consider a class of teleparallel gravity theories, whose gravitational dynamical variables are given by a Lorentzian metric \(g_{\mu\nu}\) and an affine connection with coefficients \(\Gamma^{\mu}{}_{\nu\rho}\), which is imposed to be flat,
\begin{equation}\label{eq:nocurv}
R^{\rho}{}_{\sigma\mu\nu} = \partial_{\mu}\Gamma^{\rho}{}_{\sigma\nu} - \partial_{\nu}\Gamma^{\rho}{}_{\sigma\mu} + \Gamma^{\rho}{}_{\lambda\mu}\Gamma^{\lambda}{}_{\sigma\nu} - \Gamma^{\rho}{}_{\lambda\nu}\Gamma^{\lambda}{}_{\sigma\mu} = 0\,,
\end{equation}
but in general has non-vanishing torsion
\begin{equation}\label{eq:torsion}
T^{\mu}{}_{\nu\rho} = \Gamma^{\mu}{}_{\rho\nu} - \Gamma^{\mu}{}_{\nu\rho}
\end{equation}
and nonmetricity
\begin{equation}\label{eq:nonmetricity}
Q_{\mu\nu\rho} = \nabla_{\mu}g_{\nu\rho} = \partial_{\mu}g_{\nu\rho} - \Gamma^{\sigma}{}_{\nu\mu}g_{\sigma\rho} - \Gamma^{\sigma}{}_{\rho\mu}g_{\nu\sigma}\,.
\end{equation}
There are different possibilities to impose the flatness condition~\eqref{eq:nocurv}: one may introduce a Lagrange multiplier in the action, or restrict the variation of the connection such that the flatness is maintained~\cite{Hohmann:2021fpr}. Here we follow the latter approach and restrict the variation of the geometric variables to
\begin{equation}
\delta g_{\mu\nu} = \varsigma_{\mu\nu}\,, \quad
\delta\Gamma^{\mu}{}_{\nu\rho} = \nabla_{\rho}\lambda^{\mu}{}_{\nu}\,,
\end{equation}
where \(\varsigma_{\mu\nu}\) is a symmetric tensor field, while \(\lambda^{\mu}{}_{\nu}\) is arbitrary. The action we consider thus has the general form
\begin{equation}
S[g, \Gamma, \psi] = S_{\text{g}}[g, \Gamma] + S_{\text{m}}[g, \Gamma, \psi]\,,
\end{equation}
where \(\psi^I\) denotes an arbitrary set of matter fields. Here \(S_{\text{g}}\) is the gravitational action, which needs to be specified to select a particular gravity theory, and \(S_{\text{m}}\) is a generic matter action, which defines the coupling between matter and gravity. It follows that, after integration by parts, the variation of the latter can be written in the form
\begin{equation}\label{eq:matactvar}
\delta S_{\text{m}} = \int_M\left(\frac{1}{2}\Theta^{\mu\nu}\varsigma_{\mu\nu} + \Xi_{\mu}{}^{\nu}\lambda^{\mu}{}_{\nu} + \Psi_I\delta\psi^I\right)\sqrt{-g}\dd^4x = \int_M\left(\frac{1}{2}\tilde{\Theta}^{\mu\nu}\varsigma_{\mu\nu} + \tilde{\Xi}_{\mu}{}^{\nu}\lambda^{\mu}{}_{\nu} + \tilde{\Psi}_I\delta\psi^I\right)\dd^4x\,,
\end{equation}
where \(\Psi_I = 0\) are the matter field equations, and we introduced the energy-momentum tensor \(\Theta_{\mu\nu}\) and the reduced hypermomentum \(\Xi_{\mu}{}^{\nu}\), as well as the corresponding tensor densities denoted with a tilde. For the gravitational part of the action, we similarly write
\begin{equation}\label{eq:metricgravactvar}
\delta S_{\text{g}} = -\int_M\left(\frac{1}{2}W^{\mu\nu}\varsigma_{\mu\nu} + Z_{\mu}{}^{\nu}\lambda^{\mu}{}_{\nu}\right)\sqrt{-g}\dd^4x = -\int_M\left(\frac{1}{2}\tilde{W}^{\mu\nu}\varsigma_{\mu\nu} + \tilde{Z}_{\mu}{}^{\nu}\lambda^{\mu}{}_{\nu}\right)\dd^4x\,,
\end{equation}
where \(W^{\mu\nu}\) and \(Z_{\mu}{}^{\nu}\) depend on the gravity theory under consideration. This allows us to read off the generic field equations
\begin{equation}
W_{\mu\nu} = \Theta_{\mu\nu}\,, \quad
Z_{\mu\nu} = \Xi_{\mu\nu}
\end{equation}
of general teleparallel gravity. Further, if we consider the particular variation
\begin{equation}
\varsigma_{\mu\nu} = \lc{\nabla}_{(\mu}\xi_{\nu)}\,, \quad
\lambda^{\mu}{}_{\nu} = \nabla_{\nu}\xi^{\mu} - T^{\mu}{}_{\nu\rho}\xi^{\rho}
\end{equation}
originating from an infinitesimal diffeomorphism generated by a vector field, where \(\lc{\nabla}\) denotes the Levi-Civita covariant derivative, we obtain the geometric identities
\begin{equation}\label{eq:bianchi}
\lc{\nabla}_{\nu}\tilde{W}_{\mu}{}^{\nu} + \nabla_{\nu}\tilde{Z}_{\mu}{}^{\nu} - T^{\rho}{}_{\mu\nu}\tilde{Z}_{\rho}{}^{\nu} - T^{\rho}{}_{\rho\nu}\tilde{Z}_{\mu}{}^{\nu} = 0
\end{equation}
for any diffeomorphism-invariant gravitational action, as well as the energy-momentum-hypermomentum conservation relations
\begin{equation}\label{eq:enmomhypcons}
\lc{\nabla}_{\nu}\tilde{\Theta}_{\mu}{}^{\nu} + \nabla_{\nu}\tilde{\Xi}_{\mu}{}^{\nu} - T^{\rho}{}_{\mu\nu}\tilde{\Xi}_{\rho}{}^{\nu} - T^{\rho}{}_{\rho\nu}\tilde{\Xi}_{\mu}{}^{\nu} = 0\,,
\end{equation}
which hold only on shell, i.e., when the matter field equations are satisfied~\cite{Hohmann:2022mlc}.

In the following, we are interested in the case that the dynamical fields, and hence also the field equations composed thereof, are homogeneous and isotropic. It follows that using coordinates \((\eta, r, \vartheta, \varphi)\) with conformal time \(\eta\) the metric reads
\begin{equation}
g_{\mu\nu} = -n_{\mu}n_{\nu} + h_{\mu\nu}\,,
\end{equation}
where the hypersurface conormal and spatial metric
\begin{equation}
n_{\mu}\dd x^{\mu} = -A\dd\eta\,, \quad
h_{\mu\nu}\dd x^{\mu} \otimes \dd x^{\nu} = A^2\left[\frac{\dd r \otimes \dd r}{1 - (ur)^2} + r^2(\dd\vartheta \otimes \dd\vartheta + \sin^2\vartheta\dd\varphi \otimes \dd\varphi)\right]\,,
\end{equation}
depend on the cosmological scale factor \(A = A(t)\) and (real or imaginary) curvature parameter \(u\). The connection is determined in terms of its torsion and nonmetricity as
\begin{equation}
T^{\mu}{}_{\nu\rho} = \frac{2}{A}(\mathcal{T}_1h^{\mu}_{[\nu}n_{\rho]} + \mathcal{T}_2n_{\sigma}\varepsilon^{\sigma\mu}{}_{\nu\rho})\,, \quad
Q_{\rho\mu\nu} = \frac{2}{A}(\mathcal{Q}_1n_{\rho}n_{\mu}n_{\nu} + 2\mathcal{Q}_2n_{\rho}h_{\mu\nu} + 2\mathcal{Q}_3h_{\rho(\mu}n_{\nu)})\,,
\end{equation}
where \(\varepsilon_{\mu\nu\rho\sigma}\) is the totally antisymmetric tensor normalized such that
\begin{equation}
\varepsilon_{0123} = \sqrt{-g} = \frac{A^4r^2\sin\vartheta}{\sqrt{1 - (ur)^2}}
\end{equation}
and the five functions \(\mathcal{T}_1, \mathcal{T}_2, \mathcal{Q}_1, \mathcal{Q}_2, \mathcal{Q}_3\) are constrained by the flatness condition~\eqref{eq:nocurv} to five branches in terms of the conformal Hubble parameter
\begin{equation}
\mathcal{H} = \frac{A'}{A} = \frac{1}{A}\frac{\dd A}{\dd\eta}
\end{equation}
and two further functions \(\mathcal{K}, \mathcal{L}\) of time~\cite{Heisenberg:2022mbo}. It then further follows that the field equations take the general form
\begin{equation}
W_{\mu\nu} = \mathfrak{N}n_{\mu}n_{\nu} + \mathfrak{H}h_{\mu\nu}\,, \quad
Z_{\mu\nu} = \mathfrak{T}n_{\mu}n_{\nu} + \mathfrak{S}h_{\mu\nu}\,,
\end{equation}
while for the matter side it turns out to be convenient to define
\begin{equation}\label{eq:cosmomatter}
2\kappa^2A^2\Theta_{\mu\nu} = \mathcal{D}^2n_{\mu}n_{\nu} + \mathcal{P}^2h_{\mu\nu}\,, \quad
2\kappa^2A^2\Xi_{\mu\nu} = \mathcal{E}^2n_{\mu}n_{\nu} + \mathcal{R}^2h_{\mu\nu}\,,
\end{equation}
where \(\kappa^2\) is the gravitational constant to be included in the gravitational action. In this parametrization, the energy-momentum-hypermomentum conservation~\eqref{eq:enmomhypcons} reads
\begin{equation}\label{eq:cosmocons}
2\mathcal{D}\mathcal{D}' + 2\mathcal{E}\mathcal{E}' + \mathcal{Q}_1\mathcal{E}^2 + 3\mathcal{Q}_2\mathcal{R}^2 + \mathcal{H}(\mathcal{D}^2 + \mathcal{E}^2 + 3\mathcal{P}^2 + 3\mathcal{R}^2) = 0\,.
\end{equation}
In the following, we will apply these considerations to general teleparallel quadratic gravity.

\section{Dynamical system of quadratic teleparallel gravity}
We now turn our focus to a dynamical system formulation of the quadratic teleparallel cosmology defined by the action~\cite{BeltranJimenez:2019odq}
\begin{multline}
S_{\text{g}} = -\frac{1}{2\kappa^2}\int\dd^4x\sqrt{-g}(c_1Q^{\mu\nu\rho}Q_{\mu\nu\rho} + c_2Q^{\mu\nu\rho}Q_{\rho\mu\nu} + c_3Q^{\rho\mu}{}_{\mu}Q_{\rho\nu}{}^{\nu} + c_4Q^{\mu}{}_{\mu\rho}Q_{\nu}{}^{\nu\rho} + c_5Q^{\mu}{}_{\mu\rho}Q^{\rho\nu}{}_{\nu}\\
+ a_1T^{\mu\nu\rho}T_{\mu\nu\rho} + a_2T^{\mu\nu\rho}T_{\rho\nu\mu} + a_3T^{\mu}{}_{\rho\mu}T_{\nu}{}^{\rho\nu} + b_1T^{\mu\nu\rho}Q_{\nu\rho\mu} + b_2T^{\mu\rho}{}_{\mu}Q_{\rho\nu}{}^{\nu} + b_3T^{\mu\rho}{}_{\mu}Q^{\nu}{}_{\nu\rho})
\end{multline}
parametrized by 11 constant coefficients \(a_{1,\ldots,3}, b_{1,\ldots,3}, c_{1,\ldots,5}\). The starting point for our derivation are the cosmological field equations~\cite[eq. (98-102)]{Heisenberg:2022mbo}, which we do not repeat here for brevity. These only depend on the linear combinations
\begin{gather}
z_1 = -\frac{2a_1 + a_2 + 3a_3}{2}\,, \quad
z_2 = \frac{3}{2}(3a_2 + a_3 - 2a_1)\,, \quad
z_7 = 2c_3 + c_5\,,\nonumber\\
z_3 = c_2 - c_4 + c_5\,, \quad
z_5 = b_1 + 3b_2 - 4c_1 - 12c_3\,, \quad
z_6 = b_2 + b_3 - 4c_3 - 2c_5\,,\nonumber\\
z_4 = 2a_1 + a_2 + 3a_3 - 2b_1 - 6b_2 + 4c_1 + 12c_3\,, \quad
z_8 = c_1 + c_2 + c_3 + c_4 + c_5\,.
\end{gather}
In order to obtain the desired form of a homogeneous dynamical system, we now perform the following transformations:
\begin{enumerate}
\item
To obtain a system of first-order differential equations, we eliminate the second-order derivative \(\mathcal{L}''\) by introducing a new dynamical variable \(\mathcal{M}\) as \(\mathcal{L}' = \mathcal{L}\mathcal{M}\).
\item
We replace the constant curvature parameter \(u\) by a new dynamical variable \(\mathcal{U}\), which we impose to be constant by imposing \(\mathcal{U}' = 0\).
\item
For the matter side, we use the dynamical quantities \(\mathcal{D}, \mathcal{P}, \mathcal{E}, \mathcal{R}\) introduced in~\eqref{eq:cosmomatter}.
\end{enumerate}
Applying these transformations yields for the branch 2a
\begin{subequations}
\begin{align}
\mathcal{D}^2 &= -6 (z_6+2 z_7) \mathcal{H}'-6 z_6 \mathcal{K}' + 6 (z_1-2 (z_6+2 z_7)) \mathcal{H}^2+6 (z_1+z_2) \mathcal{U}^2-8 z_8 \mathcal{M} \mathcal{L}\nonumber\\
&\phantom{=}-4 z_8 \mathcal{L}^2-6 (z_4+z_5+2 z_6) \mathcal{H} \mathcal{K}-3 z_4 \mathcal{K}^2 - 2 (3 z_6+6 z_7+8 z_8) \mathcal{H}\mathcal{L} + 6 z_6 \mathcal{K}\mathcal{L}\,,\\
\mathcal{P}^2 &= -2 (2 z_1+z_4+z_5) \mathcal{H}'+2 z_5 \mathcal{K}'-2 (z_1+2 (z_4+z_5)) \mathcal{H}^2-2 (z_1+z_2) \mathcal{U}^2+4 z_7 \mathcal{M} \mathcal{L}\nonumber\\
&\phantom{=}-4 z_8 \mathcal{L}^2-2 (3 z_4+z_5) \mathcal{H} \mathcal{K}-3 z_4 \mathcal{K}^2+ (-6 z_6-4 z_7) \mathcal{H}\mathcal{L}-6 z_6 \mathcal{K}\mathcal{L}\,,\\
\mathcal{E}^2 &= 6 (z_6+2 z_7) (\mathcal{H}' + 2\mathcal{H}^2) +6 z_6 (\mathcal{K}' + 2\mathcal{H}\mathcal{K})+16 z_8 \mathcal{H} \mathcal{L}+8 z_8 \mathcal{M} \mathcal{L}\,,\\
\mathcal{R}^2 &= 2 (z_4+z_5) (\mathcal{H}' + 2\mathcal{H}^2)+2 z_4 (\mathcal{K}' + 2\mathcal{H}\mathcal{K}) +4 z_6 \mathcal{H} \mathcal{L}+2 z_6 \mathcal{M} \mathcal{L}\,,
\end{align}
\end{subequations}
for the branch 2b
\begin{subequations}
\begin{align}
\mathcal{D}^2 &= -6 (z_6+2 z_7) \mathcal{H}'+8 z_8 \mathcal{M}'+(-6 z_6+8 z_8) \mathcal{K}' + 6 (z_1-2 (z_6+2 z_7)) \mathcal{H}^2\nonumber\\
&\phantom{=}+3 (2 z_1-2 z_3+z_4+z_5+3 z_6+4 (z_7+z_8)) \mathcal{U}^2-4 z_8 \mathcal{M}^2-9 (z_3+z_7-2 z_8)\frac{\mathcal{U}^4}{\mathcal{L}^2}\nonumber\\
&\phantom{=}-3 (-z_3+z_4+z_5+3 z_6+5 z_7+2 z_8) \mathcal{L}^2+(-6 (z_4+z_5+z_6-2 z_7)+16 z_8) \mathcal{H} \mathcal{K}\nonumber\\
&\phantom{=}+(-3 z_4+6 z_6-4 z_8) \mathcal{K}^2+ 2 (3 z_6+6 z_7+8 z_8) \mathcal{H}\mathcal{M}+(6 z_6-8 z_8) \mathcal{K}\mathcal{M}\nonumber\\
&\phantom{=}+ 3 (-4 z_3+2 z_4+3 z_5+7 z_6+18 z_7+4 z_8) \mathcal{H}\mathcal{L}+3 (2 z_4+z_5+z_6-2 z_7-4 z_8) \mathcal{K}\mathcal{L}\nonumber\\
&\phantom{=}+3 (-4 z_3+z_5-3 z_6-2 z_7+4 z_8)\frac{\mathcal{H} \mathcal{U}^2}{\mathcal{L}}+3 (z_5-3 z_6-2 z_7+4 z_8)\frac{\mathcal{U}^2 \mathcal{K}}{\mathcal{L}}\,,\\
\mathcal{P}^2 &= -2 (2 z_1+z_4+z_5) \mathcal{H}'-4 z_7 \mathcal{M}'+2 (z_5-2 z_7) \mathcal{K}' - 2 (z_1+2 (z_4+z_5)) \mathcal{H}^2\nonumber\\
&\phantom{=}-(2 z_1-2 z_3+z_4+z_5+3 z_6+4 (z_7+z_8)) \mathcal{U}^2-4 z_8 \mathcal{M}^2-(z_3+z_7-2 z_8) \frac{\mathcal{U}^4}{\mathcal{L}^2}\nonumber\\
&\phantom{=}-3 (-z_3+z_4+z_5+3 z_6+5 z_7+2 z_8) \mathcal{L}^2+(-6 z_4-2 z_5+6 z_6+4 z_7) \mathcal{H} \mathcal{K}\nonumber\\
&\phantom{=}+(-3 z_4+6 z_6-4 z_8) \mathcal{K}^2+ (6 z_6+4 z_7) \mathcal{H}\mathcal{M}+4 z_3 \frac{\mathcal{U}^2\mathcal{M}}{\mathcal{L}}\nonumber\\
&\phantom{=}-2 (-2 z_3+z_5+3 z_6+10 z_7+4 z_8) \mathcal{L}\mathcal{M}+(6 z_6-8 z_8) \mathcal{K}\mathcal{M}\nonumber\\
&\phantom{=}+ (-4 z_3+6 z_4+5 z_5+9 z_6+14 z_7-4 z_8) \mathcal{H}\mathcal{L}+3 (2 z_4+z_5+z_6-2 z_7-4 z_8) \mathcal{K}\mathcal{L}\nonumber\\
&\phantom{=}-(4 z_3+z_5-3 z_6-2 z_7+4 z_8) \frac{\mathcal{H} \mathcal{U}^2}{\mathcal{L}}-(z_5-3 z_6-2 z_7+4 z_8) \frac{\mathcal{U}^2 \mathcal{K}}{\mathcal{L}}\,,\\
\mathcal{E}^2 &= 6 (z_6+2 z_7) (\mathcal{H}' + 2\mathcal{H}^2)-8 z_8 (\mathcal{M}' + 2\mathcal{H}\mathcal{M})+2(3 z_6-4 z_8) (\mathcal{K}' + 2\mathcal{H}\mathcal{K})\nonumber\\
&\phantom{=}+6 (-z_3+z_4+z_5+3 z_6+5 z_7+2 z_8) \mathcal{L}^2+6 (z_3+z_7-2 z_8) \frac{\mathcal{U}^4}{\mathcal{L}^2}\nonumber\\
&\phantom{=}-3 (-4 z_3+2 z_4+3 z_5+7 z_6+18 z_7+4 z_8) \mathcal{H}\mathcal{L}-3 (2 z_4+z_5+z_6-2 z_7-4 z_8) \mathcal{K}\mathcal{L}\nonumber\\
&\phantom{=}+3 (4 z_3-z_5+3 z_6+2 z_7-4 z_8) \frac{\mathcal{H} \mathcal{U}^2}{\mathcal{L}}-3 (z_5-3 z_6-2 z_7+4 z_8) \frac{\mathcal{U}^2 \mathcal{K}}{\mathcal{L}}\,,\\
\mathcal{R}^2 &= 2 (z_4+z_5)(\mathcal{H}' + 2\mathcal{H}^2) -2 z_6 (\mathcal{M}' + 2\mathcal{H}\mathcal{M}) +2 (z_4-z_6) (\mathcal{K}' + 2\mathcal{H}\mathcal{K})\nonumber\\
&\phantom{=}+(4 z_3+z_5-3 z_6-2 z_7+4 z_8) \frac{\mathcal{H} \mathcal{U}^2}{\mathcal{L}}-(z_5-3 z_6-2 z_7+4 z_8) \frac{\mathcal{U}^2 \mathcal{K}}{\mathcal{L}} +2 (z_3+z_7-2 z_8) \frac{\mathcal{U}^4}{\mathcal{L}^2}\nonumber\\
&\phantom{=}+(-2 z_4-z_5-z_6+2 z_7+4 z_8) (\mathcal{K}\mathcal{L} + \mathcal{L}\mathcal{M})+ (4 z_3-6 z_4-5 z_5-9 z_6-14 z_7+4 z_8) \mathcal{H}\mathcal{L}\nonumber\\
&\phantom{=}+2 (-z_3+z_4+z_5+3 z_6+5 z_7+2 z_8) \mathcal{L}^2-(z_5-3 z_6-2 z_7+4 z_8) \frac{\mathcal{U}^2\mathcal{M}}{\mathcal{L}}\,,
\end{align}
\end{subequations}
for the branch 1a
\begin{subequations}
\begin{align}
\mathcal{D}^2 &= -6 (z_6+2 z_7) \mathcal{H}'-6 z_6 \mathcal{K}' + 6 (z_1-2 (z_6+2 z_7)) \mathcal{H}^2-8 z_8 \mathcal{M} \mathcal{L}-4 z_8 \mathcal{L}^2\nonumber\\
&\phantom{=}-6 (z_4+z_5+2 z_6) \mathcal{H} \mathcal{K}-3 z_4 \mathcal{K}^2-2 (3 z_6+6 z_7+8 z_8) \mathcal{H}\mathcal{L}-6 z_6 \mathcal{K}\mathcal{L}\,,\\
\mathcal{P}^2 &= -2 (2 z_1+z_4+z_5) \mathcal{H}'+2 z_5 \mathcal{K}'-2 (z_1+2 (z_4+z_5)) \mathcal{H}^2+4 z_7 \mathcal{M} \mathcal{L}-4 z_8 \mathcal{L}^2\nonumber\\
&\phantom{=}-2 (3 z_4+z_5) \mathcal{H} \mathcal{K}-3 z_4 \mathcal{K}^2- (6 z_6+4 z_7) \mathcal{H}\mathcal{L}-6 z_6 \mathcal{K}\mathcal{L}\,,\\
\mathcal{E}^2 &= 6 (z_6+2 z_7) (\mathcal{H}' + 2\mathcal{H}^2) +6 z_6 (\mathcal{K}' + 2\mathcal{H}\mathcal{K})+16 z_8 \mathcal{H} \mathcal{L}+8 z_8 \mathcal{M} \mathcal{L}\,,\\
\mathcal{R}^2 &= 2 (z_4+z_5) (\mathcal{H}' + 2\mathcal{H}^2)+2 z_4 (\mathcal{K}' + 2\mathcal{H}\mathcal{K}) +4 z_6 \mathcal{H} \mathcal{L}+2 z_6 \mathcal{M} \mathcal{L}\,,
\end{align}
\end{subequations}
for the branch 1b
\begin{subequations}
\begin{align}
\mathcal{D}^2 &= -6 (z_6+2 z_7) \mathcal{H}'-8 z_8 \mathcal{M}'+(-6 z_6+8 z_8) \mathcal{K}' + 6 (z_1-2 (z_6+2 z_7)) \mathcal{H}^2-4 z_8 \mathcal{M}^2\nonumber\\
&\phantom{=}-9 (z_3+z_7-2 z_8) \mathcal{L}^2+(-6 (z_4+z_5+z_6-2 z_7)+16 z_8) \mathcal{H} \mathcal{K}+(-3 z_4+6 z_6-4 z_8) \mathcal{K}^2\nonumber\\
&\phantom{=}+ 3 (-4 z_3+z_5-3 z_6-2 z_7+4 z_8) \mathcal{H}\mathcal{L}+3 (z_5-3 z_6-2 z_7+4 z_8) \mathcal{K}\mathcal{L}\nonumber\\
&\phantom{=}-2 (3 z_6+6 z_7+8 z_8) \mathcal{H}\mathcal{M}+(-6 z_6+8 z_8) \mathcal{K}\mathcal{M}\,,\\
\mathcal{P}^2 &= -2 (2 z_1+z_4+z_5) \mathcal{H}'+4 z_7 \mathcal{M}'+2 (z_5-2 z_7) \mathcal{K}'-2 (z_1+2 (z_4+z_5)) \mathcal{H}^2-4 z_8 \mathcal{M}^2\nonumber\\
&\phantom{=}+(-z_3-z_7+2 z_8) \mathcal{L}^2+(-6 z_4-2 z_5+6 z_6+4 z_7) \mathcal{H} \mathcal{K}+(-3 z_4+6 z_6-4 z_8) \mathcal{K}^2\nonumber\\
&\phantom{=}+ (-4 z_3-z_5+3 z_6+2 z_7-4 z_8) \mathcal{H}\mathcal{L}+(-z_5+3 z_6+2 z_7-4 z_8) \mathcal{K}\mathcal{L}\nonumber\\
&\phantom{=}-(6 z_6+4 z_7) \mathcal{H}\mathcal{M}-4 z_3 \mathcal{L}\mathcal{M}+(-6 z_6+8 z_8) \mathcal{K}\mathcal{M}\,,\\
\mathcal{E}^2 &= 6 (z_6+2 z_7) (\mathcal{H}' + 2\mathcal{H}^2)+8 z_8 (\mathcal{M}' + 2\mathcal{H}\mathcal{M}) + 3 (4 z_3-z_5+3 z_6+2 z_7-4 z_8) \mathcal{H}\mathcal{L}\nonumber\\
&\phantom{=}+2(3 z_6-4 z_8) (\mathcal{K}' + 2\mathcal{H}\mathcal{K})+6 (z_3+z_7-2 z_8) \mathcal{L}^2-3 (z_5-3 z_6-2 z_7+4 z_8) \mathcal{K}\mathcal{L}\,,\\
\mathcal{R}^2 &= 2 (z_4+z_5) (\mathcal{H}' + 2\mathcal{H}^2) +2 z_6 (\mathcal{M}' + 2\mathcal{H}\mathcal{M}) +2 (z_4-z_6) (\mathcal{K}' + 2\mathcal{H}\mathcal{K})\nonumber\\
&\phantom{=}+(z_5-3 z_6-2 z_7+4 z_8) \mathcal{L}\mathcal{M}+2 (z_3+z_7-2 z_8) \mathcal{L}^2\nonumber\\
&\phantom{=}+ (4 z_3+z_5-3 z_6-2 z_7+4 z_8) \mathcal{H}\mathcal{L}+(-z_5+3 z_6+2 z_7-4 z_8) \mathcal{K}\mathcal{L}\,,
\end{align}
\end{subequations}
and finally for the branch 1c
\begin{subequations}
\begin{align}
\mathcal{D}^2 &= -6 (z_6+2 z_7) \mathcal{H}'+8 z_8 \mathcal{M}'+(-6 z_6+8 z_8) \mathcal{K}' - 6 (z_1-2 (z_6+2 z_7)) \mathcal{H}^2-4 z_8 \mathcal{M}^2\nonumber\\
&\phantom{=}-3 (-z_3+z_4+z_5+3 z_6+5 z_7+2 z_8) \mathcal{L}^2+(-6 (z_4+z_5+z_6-2 z_7)+16 z_8) \mathcal{H} \mathcal{K}\nonumber\\
&\phantom{=}+ 3 (-4 z_3+2 z_4+3 z_5+7 z_6+18 z_7+4 z_8) \mathcal{H}\mathcal{L}+3 (2 z_4+z_5+z_6-2 z_7-4 z_8) \mathcal{K}\mathcal{L}\nonumber\\
&\phantom{=}+(-3 z_4+6 z_6-4 z_8) \mathcal{K}^2+ 2 (3 z_6+6 z_7+8 z_8) \mathcal{H}\mathcal{M}+(6 z_6-8 z_8) \mathcal{K}\mathcal{M}\,,\\
\mathcal{P}^2 &= -2 (2 z_1+z_4+z_5) \mathcal{H}'-4 z_7 \mathcal{M}'+2 (z_5-2 z_7) \mathcal{K}'-2 (z_1+2 (z_4+z_5)) \mathcal{H}^2-4 z_8 \mathcal{M}^2\nonumber\\
&\phantom{=}-3 (-z_3+z_4+z_5+3 z_6+5 z_7+2 z_8) \mathcal{L}^2+(-6 z_4-2 z_5+6 z_6+4 z_7) \mathcal{H} \mathcal{K}+(6 z_6-8 z_8) \mathcal{K}\mathcal{M}\nonumber\\
&\phantom{=}+(-3 z_4+6 z_6-4 z_8) \mathcal{K}^2+ (6 z_6+4 z_7) \mathcal{H}\mathcal{M}-2 (-2 z_3+z_5+3 z_6+10 z_7+4 z_8) \mathcal{L}\mathcal{M}\nonumber\\
&\phantom{=}- (4 z_3-6 z_4-5 z_5-9 z_6-14 z_7+4 z_8) \mathcal{H}\mathcal{L}+3 (2 z_4+z_5+z_6-2 z_7-4 z_8) \mathcal{K}\mathcal{L}\,,\\
\mathcal{E}^2 &= 6 (z_6+2 z_7) (\mathcal{H}' + 2\mathcal{H}^2)-8 z_8 (\mathcal{M}' + 2\mathcal{H}\mathcal{M})+2(3 z_6-4 z_8) (\mathcal{K}' + 2\mathcal{H}\mathcal{K})\nonumber\\
&\phantom{=}+6 (-z_3+z_4+z_5+3 z_6+5 z_7+2 z_8) \mathcal{L}^2-3 (2 z_4+z_5+z_6-2 z_7-4 z_8) \mathcal{K}\mathcal{L}\nonumber\\
&\phantom{=}-3 (-4 z_3+2 z_4+3 z_5+7 z_6+18 z_7+4 z_8) \mathcal{H}\mathcal{L}\,,\\
\mathcal{R}^2 &= 2 (z_4+z_5) (\mathcal{H}' + 2\mathcal{H}^2)-2 z_6 (\mathcal{M}' + 2\mathcal{H}\mathcal{M}) +2 (z_4-z_6) (\mathcal{K}' + 2\mathcal{H}\mathcal{K})\nonumber\\
&\phantom{=}+(-2 z_4-z_5-z_6+2 z_7+4 z_8) \mathcal{L}\mathcal{M} +2 (-z_3+z_4+z_5+3 z_6+5 z_7+2 z_8) \mathcal{L}^2\nonumber\\
&\phantom{=}+ (4 z_3-6 z_4-5 z_5-9 z_6-14 z_7+4 z_8) \mathcal{H}\mathcal{L}+(-2 z_4-z_5-z_6+2 z_7+4 z_8) \mathcal{K}\mathcal{L}\,.
\end{align}
\end{subequations}
Due to the large number of undetermined parameters, we restrict our discussion to the generic properties of these systems of equations, and defer a more detailed analysis to future work. Denoting the set of dynamical variables by
\begin{equation}
\boldsymbol{\mathcal{X}} = (\mathcal{X}^I) = (\mathcal{H}, \mathcal{K}, \mathcal{L}, \mathcal{M}, \mathcal{U}, \mathcal{D}, \mathcal{P}, \mathcal{E}, \mathcal{R})\,,
\end{equation}
the field equations, the conservation equation~\eqref{eq:cosmocons} and the definitions of \(\mathcal{U}'\) and \(\mathcal{L}'\) read
\begin{equation}
C_{IJ}{\mathcal{X}^J}' + f_I(\boldsymbol{\mathcal{X}}) = 0\,,
\end{equation}
where \(C\) is a constant, rectangular matrix and
\begin{equation}
f_I(\lambda\boldsymbol{\mathcal{X}}) = \lambda^2f_I(\boldsymbol{\mathcal{X}})
\end{equation}
are homogeneous functions of order 2. First, note that this system is underdetermined in general and must be supplemented by an equation of state relating the matter variables \(\mathcal{D}, \mathcal{P}, \mathcal{E}, \mathcal{R}\). Further, one finds that the appearance of time derivatives in the resulting equations is not linearly independent, and so one can isolate constraints between the dynamical variables, i.e., equations without time derivatives. Denoting the remaining free variables, after imposing constraints, as \(\boldsymbol{\mathcal{Y}} = (\mathcal{Y}^i)\), we are left with the system
\begin{equation}
B_{ij}{\mathcal{Y}^j}' + F_i(\boldsymbol{\mathcal{Y}}) = 0\,,
\end{equation}
where \(B\) is now a constant, quadratic, invertible matrix and
\begin{equation}
F_i(\lambda\boldsymbol{\mathcal{Y}}) = \lambda^2F_i(\boldsymbol{\mathcal{Y}})\,,
\end{equation}
so that we can solve the system as
\begin{equation}
{\mathcal{Y}^i}' = -(B^{-1})^{ij}F_j(\boldsymbol{\mathcal{Y}}) = Z^i(\boldsymbol{\mathcal{Y}})\,.
\end{equation}
At this point it helps to introduce homogeneous variables
\begin{equation}
Y = \|\boldsymbol{\mathcal{Y}}\| = \sqrt{\delta_{ij}\mathcal{Y}^i\mathcal{Y}^j}\,, \quad
N^i = \frac{\mathcal{Y}^i}{Y}\,, \quad
\mathbf{N} = (N^i)\,.
\end{equation}
We can thus write the dynamical system as
\begin{equation}
{\mathcal{Y}^i}' = Y'N^i + Y{N^i}' = Y^2Z^i(\mathbf{N})\,,
\end{equation}
where
\begin{equation}
Y' = Y^2\delta_{ij}Z^i(\mathbf{N})N^j\,, \quad
{N^i}' = Y[Z^i(\mathbf{N}) - \delta_{jk}Z^j(\mathbf{N})N^kN^i]\,.
\end{equation}
Finally, introducing a new time coordinate by \(\dd T = Y\dd\eta\), and writing \(\dot{ } = \dd/\dd T\), we have
\begin{equation}
\dot{Y} = Y\delta_{ij}Z^i(\mathbf{N})N^j\,, \quad
\dot{N}^i = Z^i(\mathbf{N}) - \delta_{jk}Z^j(\mathbf{N})N^kN^i\,.
\end{equation}
Note that the second equation does not depend on \(Y\). We call points with \(\dot{N}^i = 0\) \emph{projective fixed points}. It follows from the fact that \(\mathbf{N}\) is normalized, and thus belongs to the compact unit sphere, that such points always exist. At these points, the radial dynamics satisfies
\begin{equation}
Y = Y_0\exp[\delta_{ij}Z^i(\mathbf{N})N^jT] = \frac{1}{\delta_{ij}Z^i(\mathbf{N})N^j(\eta_0 - \eta)}\,,
\end{equation}
showing the presence of a future or past singularity at \(\eta = \eta_0\), accompanied by an asymptotic \(Y \xrightarrow{\eta \to \pm\infty} 0\). The exact behavior depends on the theory under consideration and must be determined along the lines of~\cite{Hohmann:2025kxp}.

\section{Conclusion}
We discussed the cosmological dynamics of general teleparallel quadratic gravity. Using a suitable set of dynamical variables, we expressed the cosmological field equations as a dynamical system, and showed that the time derivatives of the dynamical variables are given by two-homogeneous functions. We used this fact to separate these equations into angular and radial parts. From the former, we found the existence of projective fixed points, where the dynamics of the system becomes purely radial. We further solved the radial dynamics at these points and found that it exhibits future or past finite time singularities as well as asymptotically vanishing quantities such as the conformal Hubble parameter.

Our results are in line with similar results in the case of Newer General Relativity, which is a related theory based on symmetric teleparallel gravity~\cite{Hohmann:2025kxp}. Due to the complexity of the general teleparallel gravity theory, however, we restricted ourselves to a study of generic features in this work. A full discussion of the theory is needed in order to determine the number, location and stability of projective fixed points in the cosmological phase space, along with their resulting radial dynamics. For this task to become feasible, constraining the cosmological parameters \(z_{1, \ldots, 8}\) by other phenomenology, such as gravitational waves or the parametrized post-Newtonian limit, is an appropriate next step. We therefore leave a detailed analysis for future investigation.

\begin{acknowledgments}
The author thanks Remo Ruffini and Gregory Vereshchagin for the kind invitation to participate in the Sixth Zeldovich Meeting.
\end{acknowledgments}

\section*{Funding}
The author gratefully acknowledges the full financial support by the Estonian Research Council through the Personal Research Funding project PRG2608 and the Center of Excellence TK202 ``Fundamental Universe''. Participation in the Sixth Zeldovich Meeting was supported by the COST action CA21136 CosmoVerse.

\bibliographystyle{aipnum4-2}
\bibliography{references}
\end{document}